%Paper: hep-th/9303147
%From: smassar@ulb.ac.be
%Date: Fri, 26 Mar 1993 12:13:33 +0100

%
%  goodies for ref's
%

\def\RREF#1#2{\gdef#1{\REF#1{#2}#1}}	%%%%%%%%%%%%%%Relocatable \REF - a try

\def\jnl#1&#2(#3){\begingroup\let\jnl=\dummyj@urnal\sl #1\bf#2
\rm (\afterassignment\j@ur\count255=#3)\endgroup}	%no automatic ","
%
%  RefERENCES %%%%%%%%%%%%%%%%%%%

%
%

\overfullrule=0pt

\def\sqr#1#2{{\vcenter{\vbox{\hrule height.#2pt
          \hbox{\vrule width.#2pt height#1pt \kern#1pt
           \vrule width.#2pt}
           \hrule height.#2pt}}}}

\def\square{\mathchoice\sqr68\sqr68\sqr{4.2}6\sqr{3}6}

\def\lrpartial{\mathrel{\partial\kern-.75em\raise1.75ex
\hbox{$\leftrightarrow$}}}

\RREF\hawk{S.W. Hawking, Commun. math. Phys. {\bf 43}, 199 (1975).}

\RREF\birreld{N.D. Birrel and P.C.W. Davies, Quantum Fields in Curved
Space,
Cambridge University Press (1982).}

\RREF\candelas{P. Candelas, Phys. Rev. D {\bf 21}, 2185 (1980)\nextline
K.W.Howard, Phys. Rev. D {\bf 30}, 2532 (1984)}

\RREF\dfu{P.C.W. Davies, S.A. Fulling and W.G. Unruh, Phys. Rev. D{\bf 13}
, 2720 (1976).}

\RREF\pbone  {R. Parentani and R. Brout, Int. J. Mod. Phys. D{\bf 1}
,169 (1992).}

\RREF\unruh{W.G. Unruh, Phys. Rev. D {\bf 14}, 870 (1976).}

\RREF\bunchp{T.S. Bunch and L. Parker, Phys. Rev. D {\bf 20}, 2499 (1979).}

\RREF\damour  {T. Damour and R. Ruffini, Phys. Rev. D{\bf 14}, 332 (1976).}

\RREF\pbtwo  {R. Parentani and R. Brout, Nucl. Phys. B (in press).}

\RREF\parentani  {R. Parentani, Brussels preprint in preparation.}

\RREF\punsly {After this manuscript was written we came across a similar
development by B. Punsly, Phys. Rev. D{\bf 46}, 1288 ( 1992).
This reference however does not carry out the program in
detail as we have done and so misses the thermal-like
character of the radiation at finite distance from the horizon.
Moreover, his realization of the inertial modes differs from our
more simple prescription.}

\date{January 1993}
\titlepage \line{\hfill\vbox{
\hbox{ULB-TH-01/93}}}

\vskip0.2cm plus 0.2fil
\title{ Energy Momentum Tensor of the Evaporating
Black Hole and Local Bogoljubov
Transformations}
\vskip0.5cm plus 0.7fil
\centerline{
S.Massar\foot{Boursier IISN, e-mail: ulbg062@bbrnsf11.bitnet}$^a$,
R. Parentani\foot{e-mail: Renaud@vms.huji.ac.il}$^b$,
R. Brout$^a$}
\vskip 0.5cm plus 0.1fil
\centerline{\vbox{\hsize=13cm \it \halign{#&#\hfil\cr
&a Service de Physique Theorique, Universite Libre de Bruxelles \cr
&\quad Campus Plaine, C.P. 225, Bd du Triomphe, B-1050 Brussels, Belgium
\cr &b Departement of Theoretical Physics,
The Racah Institute of Physics\cr
&\quad The Hebrew University of Jerusalem, Givat Ram Campus,
Jerusalem 91904, Israel
\cr}}}
\vskip 0.5cm plus 0.7fil
\centerline{Abstract}
The method of Hawking to obtain black hole evaporation through Bogoljubov
transformation between asymptotic modes (in and out) is generalized.  The
construction is local in that the in modes (of say positive frequency) are
decomposed by Bogoljubov transformation into positive and negative
frequency
local inertial modes (i.e. those which are solutions of the d'Alembertian
in
terms of the local normal coordinates).  From this follows an interesting
reexpression of the local energy momentum tensor, more particularly of the
outgoing energy flux.  One finds that even in the local description there
exists a partial thermal character parametrized by a local
temperature.  There exists quantum interference effects as well.  These
become
negligible at large distance from the black hole.
\endpage

\chapter {Introduction}

There exists in the literature two approaches to black hole evaporation :
\hfil\break
\hbox{\ \ \ } 1] mode by mode in the manner of Hawking in his seminal work
on the
subject\refmark\hawk;
\hfil\break
\hbox{\ \ \ } 2] globally, through calculation of the (subtracted) energy
momentum
tensor $\langle T_{\mu \nu} \rangle_{ren}$\hfil\break
\hbox{\ \ \ } of the evaporated radiation (i.e.
summed on all modes) \refmark{\birreld}\refmark{\dfu}.\hfil\break
\noindent The mode by mode calculation is generally used
to calculate $\langle T_{\mu \nu} \rangle_{ren}$ asymptotically ($r \gg
r_s$
where $r_s$ = Schwarzschild horizon radius = $2M$) whereas the global
method
yields $\langle T_{\mu \nu} \rangle_{ren}$ for finite $r$ as well.  In
reference \pbtwo ,
it was only qualitatively indicated how to get a handle on what happens
locally
by mode decomposition.  In the present work the local behavior through mode
decomposition is developed quantitatively by generalizing Hawking's
Bogoljubov
transformation as follows.

In calculating $\langle T_{\mu \nu} \rangle_{ren}$ one subtracts off
$\langle T_{\mu \nu} \rangle_0$, the (infinite) expectation value of
$T_{\mu \nu}$ in the local inertial vacuum.  One constructs normal
coordinates
around the given space-time point $P$ and quantizes fields in terms of the
modes defined by that coordinate system so as to obtain the local inertial
vacuum.  Indeed this is the state which the inertial observer at $P$
perceives
as vacuum (in the cosmological context it goes by the name of
adiabatic vacuum, but can be defined quite generally \refmark{\bunchp} ).
 Any other state (including vacua defined in terms of other bases)
is excited.  This is one way to look at Hawking's calculation wherein modes
radiated out of the collapsing star (incipient black hole) are
``bogoljubized''
into Schwarzschild modes.  Since these latter are inertial at infinity, and
the
space is flat there as well, the Bogoljubov coefficients are sufficient to
calculate  $\langle T_{\mu \nu} \rangle_{ren}$ at $r = \infty$.  We now
generalize this by ``bogoljubizing'' the outgoing modes into local inertial
modes at any $P$ \refmark{\punsly}.  In so doing we display
the build-up of the thermal
flux as $r$ increases from $r_s$ to $\infty$.  The task is facilitated by
using
Unruh's \refmark{\unruh} isomorphism wherein the true outgoing modes are
replaced by
Kruskal modes, a procedure which is valid for asymptotic Schwarzschild
times.
In this way the Bogoljubov transformation is simply the transformation from
quanta built on the Unruh vacuum to those of the local inertial vacuum.

The result of the calculation is that the contribution of the local
inertial
modes to $\langle T_{\mu \nu} \rangle$ has an (approximate) thermal
character given in terms of a local
temperature $T(r)=T_H (1-2M/r)$ where $T_H$ is the Hawking temperature
$T_H = 1/(4 \pi r_s)$. At large $r$ and large $\lambda$ the
interference terms between
Bogoljubov coefficients vanish exponentially and
at
$r=\infty$ the Planck distribution is recoverd.

The  interest of this mode by mode decomposition is its potential
usefulness
in clarifying the gravitational back reaction . Just how does the
collapsing
star (incipient black hole) lose is mass as {\bf one}
 Hawking photon is evaporated ?
For this one needs a particle (i.e. a mode) description. The tunneling
mechanisms responsible for the evaporation will then be brought
to bear \refmark{\pbone}\refmark{\pbtwo}.
We have already made some progress in this direction in the analogous, but
somewhat easier, problem of pair production in a static electric field
and expect to report on this in a subsequent publication.

Section $2$ contains the general procedure in any curved space and
section $3$ is the same in terms of modes. Section $4$ deals with
the black hole radiation.

\chapter{ The Renormalised Stress Tensor}

We shall work in $1 + 1$ dimensions.  It is well-known in the black hole
problem, because of S-wave dominance in the emission, that apart
from an irrelevant transmission coefficient, this reduction of
dimensionality does no injustice to the physics.  Furthermore, for
simplicity we shall deal with a zero mass scalar field.  Then the
calculation of $\langle T_{\mu \nu} \rangle$, thougt conserving the
characteristic
space-time dependence of the full $4$-dimensional stress-tensor
\refmark{\candelas}, is greatly facilitated owing to the conformal
flatness of two-dimensional geometry.

First, we calculate the difference $\Delta \langle T_{\mu \nu} \rangle$
between two vacuum states $\ket{\Omega_{u,v}}$ and $\ket{\Omega_{f,g}}$.
These vacua refer to modes which are quantized in different coordinate
systems (but the same geometry), which are transformed into each other
by conformal transformation: $u \rightarrow f(u)$ ; $v \rightarrow g(v)$ .

Here $(u,v)$ are light-like, so that
$$ds^2 = C(u,v)dudv=C(u,v) f^{\prime -1}  g^{\prime -1} dfdg
\eqn\eqIIi$$
where in the second part of the equality $u=u(f)$,$v=v(g)$
denote the inverse transformations.

We shall choose the indices $\mu \nu$ to be either $u$ or $v$ and
begin by calculating
$\Delta \langle T_{uu} \rangle =\bra{\Omega_{u,v}} T_{uu}
\ket{\Omega_{u,v}}
- \bra{\Omega_{f,g}} T_{uu} \ket{\Omega_{f,g}}$,
where $T_{uu} = \left ( {\partial \phi \over \partial u }\right)^2$ and
$\phi$ is the quantum field.
The modes are $e^{-i \omega u}/\sqrt{4 \pi \omega}$,
$e^{-i \omega v}/\sqrt{4 \pi \omega}$
 and $e^{-i \lambda f(u)}/\sqrt{4 \pi \lambda}$,
$e^{-i \lambda g(v)}/\sqrt{4 \pi \lambda}$ in the $(u,v)$ and $(f,g)$
systems respectively,
so that formally
$\Delta \langle T_{uu} \rangle = {1\over 4 \pi}
\left \lbrack\int_0^\infty\! \omega d \omega
- ( f^\prime (u) )^2 {1\over 4 \pi}\int_0^\infty\! \lambda d
\lambda \right \rbrack $.
This formally infinite expression must be
regularized in order to define it as a limit.  To this end one
introduces $G_+$ (positive frequency Wightman function)
$$\eqalign{G_+( u^\prime,u;v^\prime,v) & = {1\over 4 \pi}
\lbrack\int_0^\infty {d \omega \over \omega}
\lbrack e^{i \omega (u^\prime - u)} + e^{i \omega (v^\prime - v)} \rbrack
\cr
&= - {1\over 4 \pi} \left ( ln\vert u^\prime - u\vert +
ln\vert v^\prime - v\vert \right )
\equiv G_+(u^\prime , u) + G_+(v^\prime , v) \cr}\eqn\eqIIii$$
and similarly in the $(f,g)$ system.  To calculate $\langle T_{uu} \rangle$
we
need only consider $G_+(u^\prime , u)$.  One then defines the limit
$$\Delta \langle T_{uu} \rangle =  - {1\over 4 \pi}
\lim_{u^\prime - u \rightarrow 0}{ \left \{ \partial_u \partial_{u^\prime}
\left( ln\vert u^\prime - u\vert -
ln\vert f(u^\prime) - f(u) \vert \right)\right \} }\eqn\eqIIiii$$
That this difference is finite is the content of
Hadamard's theorem, which assures that the short-distance
singularities of Green functions are always the same.
Direct calculation yields
$$\Delta \langle T_{uu} \rangle = {1\over 4 \pi} \left \lbrack
{f^{\prime\prime\prime} \over 6 f \prime} -
{ f^{\prime\prime 2} \over 4 f^{\prime 2}} \right \rbrack
= -{1\over 12 \pi} f^{\prime {1\over 2}} \partial^2_u f^{\prime -{1\over
2}}
\eqn\eqIIiv$$

We apply this result to calculate the renormalized $\langle T_{\mu \nu}
\rangle$
in a given geometry.  The coordinates $u,v$ refer to a set which has been
picked for some physically interesting reason, generally related to
asymptotic boundary conditions.  From $\langle T_{\mu \nu} \rangle$
calculated in this system
we must subtract  $\langle T_{\mu \nu} \rangle_0$, calculated
in the local inertial system $(\hat u, \hat v)$.
Note that the coordinates $\hat u, \hat v$ should bear an additional label,

say $(u_0 , v_0 )$, to indicate the point about
which they are inertial.
 We shall not encumber the notation by this label, since the
special role of $(u_0 , v_0 )$ will become obvious in context.

$\hat u$ is given
by (if the metric is 2.1)
$$ \hat u - \hat u_0 =
\int^u_{u_0} {C(u^\prime , v_0) \over C(u_0 , v_0) } d u^\prime
\eqn\eqIIv$$
and similarly for $\hat v$ .  We then have
$$ds^2 = C(u,v)dudv={ C(u,v) C^2 (u_0 , v_0) \over C(u_0,v) C(u,v_0)}
d \hat u d\hat v \eqn\eqIIvi$$
where, in the second part of the equality,
it is understood that $u,v$ are functions of $\hat u , \hat v$ obtained by
inversion of 2.5).  Eq. 2.6) shows that $\hat u , \hat v$ are indeed
inertial
in that the expansion about $(u_0,v_0)$ of the conformal factor begins at
second
order in $(\hat u - \hat u_0 )$ and $(\hat v - \hat v_0 )$.
These locally inertial coordinates are the
affine parameters which locate points on the light-like
trajectories $u=u_0$, $v=v_0$ , and are thus independent of the
initial choice $(u,v)$.  Note that $\hat u$ is defined up to a
scale factor which has been taken in 2.5) such that
${d \hat u \over d u}(u=u_0)=1$.

	The renormalized $\langle T_{\mu \nu} \rangle$ is defined by
$$\langle T_{\mu \nu} \rangle_{ren} = \langle T_{\mu \nu} \rangle
-\langle T_{\mu \nu} \rangle_0 \eqn\eqIIvii$$
the difference between vacuum with respect to
modes in the $(u,v)$ coordinate system and that of
the local inertial system defined by $(\hat u , \hat v)$.
So the source for the gravitational field
(the renormalised energy momentum tensor)
is exactly that energy which a local inertial
detector would measure.  The function $f(u)$ in Eq. 2.1)
is replaced by $\hat u (u )$, which from 2.5 gives $f^\prime (u) =
C(u,v_0) / C(u_0,v_0)$.  Thus we recover the well known
result\refmark{\birreld}\refmark{\dfu}:
$$\langle T_{uu} \rangle_{ren} =
-{1 \over 12 \pi} C^{1/2} \partial^2_u C^{-1/2}
\eqn\eqIIviii$$
	A similar result obtains for $\langle T_{vv} \rangle_{ren}$.
The same technique would give $\langle T_{uv} \rangle_{ren}=0$, since
the operator $T_{uv}$ is identically zero.
But then the stress tensor, though
traceless, is not conserved.
The non-conservation stems from the parametric dependence
of the local vacuum on $u_0,v_0$ (through the $u_0,v_0$ label the inertial
coordinates $\hat u , \hat v$ bear (cf. their definition 2.5)).
The difference in energy in between two vacua (2.4)
is conserved but the local vacuum changes from point
to point and the difference in energy in between the
physical vacuum and the local vacuum (2.7) is not
conserved : $\langle T_{uu} \rangle_{ren},v \not = 0$
(in the next section the $u_0,v_0$ dependence
of the local Bogoljubov coefficients will
encode the non conservation).
The trace anomaly restores energy
conservation at the price of conformal
invariance and may be obtained by imposing
$\langle T_\nu^\mu \rangle_{ren\ ;\mu} = 0$
or by more refined limiting procedures \refmark{\dfu}.
One finds
$$\langle T_{uv} \rangle_{ren}={1 \over 96 \pi} C(u,v)R(u,v)\eqn\eqIIix$$
where $R(u,v)=\square ln C(u,v)$ is the curvature.  We shall focus on
$\langle T_{uu} \rangle_{ren}$,
since it is herein that lies the Hawking radiation.

	We review briefly how equations 2.4) and 2.8)
are applied to the black hole problem in the
manner of Davies, Fulling and Unruh \refmark{\dfu}.
Here the essential ingredient is the remark of
Unruh \refmark{\unruh}, wherein the behavior of a quantum
field in the space-time of an incipient black hole
(collapsing star at asymptotic Schwarzschild time)
are those of a quantum state: Unruh vacuum $(\equiv \ket {U} )$.
This state is vacuum with respect to outgoing $(U)$
modes which are Kruskal and incoming $(v_S)$
Schwarzschild modes.  In usual notation
$$\eqalign{ u_S \equiv t-r^\ast\qquad &;\qquad U \equiv -e^{-u_S/4M}\cr
v_S \equiv t+r^\ast\qquad &;\qquad V \equiv e^{v_S/4M}\cr}\eqn\eqIIx$$
where $(u_S,v_S)$ is the Schwarzschild set (noted throughout the paper with
a
subscript $S$ to distinguish it from the generic set
$u,v$ introduced previously) and $(U,V)$ Kruskal.

	The change induced in the modes due to collapse
is to convert $u_S$ modes to $U$ modes.  Concomitantly,
the vacuum changes from Boulware $(\equiv \ket {B} )$ to $\ket {U}$,
where $\ket {B}$
is vacuum with respect to both $u_S$ and $v_S$ modes.
Hence from 2.4) and 2.10)
$$\Delta \langle T_{u_S u_S} \rangle \equiv \bra{U}T_{u_Su_S} \ket{U}
-\bra{B}T_{u_Su_S} \ket{B} = {\pi \over 12} {1 \over (8 \pi M)^2} =
{\pi \over 12} T^2_H \eqn\eqIIxi$$
The difference is independent of $r$.
In virtue of staticity and the $u_S,v_S$ symmetry
of $\ket{B}$ vacuum, the right hand side of 2.11) is the Hawking
flux, here encoded in $\ket{U}$.  Furthermore, the
dependence of $\langle T_{\mu \nu} \rangle$ on $r$ is the same in $\ket{B}$
 and $\ket{U}$,
therefore obtainable from 2.8 with the
Schwarzschild metric ($C(r) =(1-2M/r)$).  For example,
$$\bra{U}T_{u_Su_S} \ket{U} = {\pi \over 12} {1 \over (8 \pi M)^2}
\left(1-{2M \over r}\right)^2 \left(1 + {4M \over r} +
{ 12 M^2 \over r^2}\right)\eqn\eqIIxii$$

\chapter {Mode by mode decomposition of the stress tensor}

	In Section 2 there was displayed two bases
of quantization: \hfil\break
\hbox{\ \ \ } 1]the inertial modes
$e^{-i\lambda (\hat u - \hat u_0 ) } / \sqrt{ 4 \pi \lambda} \
(\equiv \xi_\lambda)$ and
\hfil\break
\hbox{\ \ \ } 2] the modes of physical interest
$e^{-i\omega (u - u_0 ) } / \sqrt{ 4 \pi \omega} \
(\equiv \chi_\omega)$,\hfil\break
 with $u$ and
 $\hat u$ related through \hfil\break
Eq. 2.5) and where we have
tuned them to equal phase at $u_0$.\hfil\break
They are
related one to the other by a Bogoljubov transformation
$$\chi_\omega (u) = \int_0^\infty \! d \lambda \ \
\alpha_{\omega \lambda }\ \xi_\lambda (\hat u (u))
-\beta_{\omega \lambda }\ \xi_{- \lambda} (\hat u (u))
\eqn\eqIIIi$$
$$\hbox{where} \qquad  \left\{ \matrix{ \alpha_{\omega \lambda }\cr
\beta_{\omega \lambda}\cr} \right\} = i \int_{- \infty}^{+\infty} \ du
\left\{ \matrix{\xi_{- \lambda} (\hat u (u))\cr
\xi_\lambda (\hat u (u))\cr} \right\}
\lrpartial_u \chi_\omega (u) \eqn\eqIIIii$$

and one checks out unitarity in the form
$$\int_0^{+\infty}  \!\! d\omega\  \lbrack \alpha_{\omega \lambda }^\ast
\alpha_{\omega \lambda^\prime } - \beta_{\omega \lambda}
 \beta_{\omega \lambda^\prime}^\ast \rbrack = \delta(\lambda -
\lambda^\prime)
\eqn\eqIIIiii$$
(and similarly with the roles of $\omega$ and $\lambda$ reversed).

	Our project is first to express $\langle T_{uu} \rangle_{ren}$
 in terms of the $\alpha$'s
and $\beta$'s in the general case and then apply the formulae
to black hole evaporation.  Using $\langle T_{uu} \rangle
= \bra{\Omega_u} (\partial \phi / \partial u )^2 \ket{\Omega_u}$
were $\ket{\Omega_u}$ is the physical vacuum  and expressing $\chi_\omega$
in terms of $\xi_\lambda$ yields
$$\langle T_{uu} \rangle_{ren} = {1\over 4\pi}
\left ( {d \hat u \over du }\right)^2
\left \lbrack
\int_0^\infty \! d \lambda \int_0^\infty \! d \lambda^\prime
\ \sqrt{\lambda\lambda^\prime}\  \lbrack \alpha^2_{\lambda\lambda^\prime} +
\beta^2_{\lambda\lambda^\prime} + \alpha^\ast \beta_{\lambda\lambda^\prime}
+
\alpha \beta^\ast_{\lambda\lambda^\prime} \rbrack
\quad - \quad \int_0^\infty\lambda d \lambda \right \rbrack
\eqn\eqIIIiv$$
where
$$\alpha^2_{\lambda\lambda^\prime} = \int_0^\infty d \omega
\alpha_{\omega \lambda }\alpha_{\omega \lambda^\prime }^\ast
\eqn\eqIIIv$$
and similarly for $\beta^2_{\lambda\lambda^\prime}$,
$\alpha^\ast \beta_{\lambda\lambda^\prime}$,
$\alpha \beta^\ast_{\lambda\lambda^\prime}$. Notice that in 3.4) the
$\alpha$
and $\beta$ coefficients depend on $v_0$ (cf. Eq. 2.5)) so
that the distribution of local particles changes with $v$ along
a geodesic $u=u_0$.
As discussed in Section $2$ the $v$ dependence of the bogoljubov
coeficients
renders $\langle T_{uu} \rangle_{ren\ ,v} \not = 0$, which would result
in non conservation were it not for the anomaly.

	Before proceeding to the black hole problem, it is
instructive to see how 3.4) leads back to 2.8) as an
identity.  Putting the explicit forms 3.2) into 3.4)
gives
$$\eqalign{ \langle T_{uu}(u_0,v_0) \rangle_{ren}& =
\int^{+ \infty}_{- \infty} \! d \lambda
\int^{+ \infty}_{- \infty} \! d \lambda^\prime \int^{+ \infty}_0 \! d
\omega
\int \! d \hat u_1 \int \! d \hat u_2
{\sqrt{\lambda \lambda^\prime} \over 4 \pi} \ \times \cr
& \qquad\times {e^{+i\lbrack \lambda ( \hat u_1 - \hat u_0 ) -
\lambda^\prime ( \hat u_2 - \hat u_0 ) \rbrack} \over
 4 \pi\sqrt{\lambda \lambda^\prime}  }
\ (i \lrpartial_{\hat u_1} ) (-i \lrpartial_{\hat u_2} )\
{e^{-i\omega \lbrack u(\hat u_1) - u(\hat u_2) -i \epsilon
\rbrack} \over 4 \pi \omega}\cr
&-\int^{+ \infty}_0 {\lambda d \lambda \over 4 \pi}\cr}
\eqn\eqIIIvi$$
Some interesting features have been exploited in
arriving at this expression.  Firstly, the formulae 3.2)
 have been rewritten with $u$ and $\hat u$ interchanged and the
function $u(\hat u)$, the inverse of 2.5), has been introduced.
Secondly the jacobian factor
$\left ( {d \hat u \over du }\right)^2$ in 3.4) has been droped
since, as follows from 2.5), it is equal to $1$ at the point of
interest $(u_0,v_0)$.
Thirdly, it will be noted that the $\lambda$ integrations in 3.6)
go over negative and positive values whereas in 3.4)
they are positive only.  This rewriting follows from
the definitions 3.2), wherein the $\alpha$ ($\beta	$) are defined for
negative (positive) $\lambda$ respectively when $\omega$ is positive.
Regrouping all the terms in 3.4) then gives the simpler
expression 3.6).
Performing the integrations over $\lambda$, $\lambda^\prime$,
$\omega$
gives the singular form
$$\eqalign{ \langle T_{uu} (u_0,v_0)\rangle_{ren}& =
\!\int \! d \hat u_1\! \int \! d \hat u_2 \!
\left \{ \partial_{\hat u_1}\partial_{\hat u_2}
{(-1) \over 4 \pi} ln \vert u(\hat u_1) - u(\hat u_2) -i \epsilon \vert
\right \} \delta ( \hat u_1 - \hat u_0 )
\delta ( \hat u_2 - \hat u_0 ) \cr
&\qquad - \int^{+ \infty}_0\! {\lambda d \lambda \over 4 \pi}\cr}
\eqn\eqIIIvii$$
In the integrand of the first integral in 3.7),
one then takes the derivatives and expands all
functions around $\hat u_0 $.  There results a divergent
term and a convergent remainder.  The former is
easily shown to be equal and opposite to the
subtraction when this latter is written in
terms of the split point regulator.  The finite
remainder is
$$\langle T_{uu}(u_0,v_0) \rangle_{ren}
= - {1 \over 4 \pi} \left \lbrack { u^{\prime\prime\prime} \over 6
u^\prime}
-{ u^{\prime\prime 2} \over 4 u^{\prime 2}} \right \rbrack
\eqn\eqIIIviii$$
where prime means derivative with respect to $\hat u$.
This is easily checked to be a rewrite of 2.8)
in terms of the inverse function $u(\hat u)$.  It is amusing
that the algebra in the reduction of 3.7) to 3.8) is
identical to the passage from 2.3) to 2.4).

	Interesting physics is revealed by dissecting
Eq. 3.6) into its positive and negative $\lambda$ pieces
as in 3.4).  To this end we record the following
formula for the Bogoljubov matrices
$\alpha^2_{\lambda\lambda^\prime}$, $\beta^2_{\lambda\lambda^\prime}$,
$\alpha^\ast \beta_{\lambda\lambda^\prime}$,
$\alpha \beta^\ast_{\lambda\lambda^\prime}$
 defined
in 3.5).  They are obtained by direct substitution
of 3.2), judicious use of integration by parts to
get symmetric forms, integration over $\omega$ and passage
to the variables $x=(\hat u_1 +\hat u_2)/2$ ; $\Delta = \hat u_1 -\hat u_2$
 where $\hat u_1$,$\hat u_2$ are the integration
variables in the definitions 3.2) (obtained after change
of variable of integrations from $u$ to $\hat u$).
$$\eqalign{\left \{ \matrix {
\alpha^2_{\lambda\lambda^\prime}\cr
\beta^2_{\lambda\lambda^\prime}\cr
\alpha^\ast \beta_{\lambda\lambda^\prime}\cr
\alpha \beta^\ast_{\lambda\lambda^\prime}\cr}\right\}
&= \!\int \!dx \!\int \!d \Delta {i \over 8 \pi^2
\sqrt{\lambda\lambda^\prime} }
e^{+i(\lambda - \lambda^\prime)
(x-\hat u_0) } \
e^{+i({\lambda + \lambda^\prime \over 2})
\Delta } \ \times \cr
& \times \ \left \lbrack
{ ({\lambda + \lambda^\prime \over 2})
\left(u^\prime ( x+ {\Delta\over 2}) +
u^\prime ( x- {\Delta\over 2}) \right)
\ +\ ({\lambda - \lambda^\prime \over 2}
)\left(u^\prime ( x+ {\Delta\over 2}) -
u^\prime ( x- {\Delta\over 2}) \right)
\over u ( x+ {\Delta\over 2}) -
u ( x- {\Delta\over 2}) - i \epsilon }
 \right \rbrack \cr}\eqn\eqIIIix$$
In $\alpha^2$,$\lambda$ and $\lambda^\prime$ are both positive;
in $\beta^2$ they are both negative
and in $\alpha^\ast \beta$ and $\alpha \beta^\ast$
 of opposite signs.  Assume the
function $u(\hat u)$ such as to permit contour
integration by closing the $\Delta$ integration on the
half circle of $\infty$ radius (confirmed to be
legitimate in the black hole case).  Let $\Delta_n(x)$ be
the poles, i.e.,
$$u(x + \Delta_n /2) - u(x - \Delta_n /2) -i\epsilon =0
\eqn\eqIIIx$$
There is one pole at $\Delta_0 = +i\epsilon$.  Reality and symmetry
prescribe that other poles come in either
doublets or quartets: $\Delta_n = \pm A_n \pm i B_n$.  This doubling of all

poles in the upper and lower planes, but only
one pole at $i\epsilon$ is a reflection of the unitarity
condition 3.3).  Then 3.9) becomes (using the
relation $\left \lbrack u^\prime (x + \Delta_n /2) -
u^\prime (x - \Delta_n /2)\right \rbrack
+ {\Delta^\prime_n \over 2}\left \lbrack u^\prime (x + \Delta_n /2) +
u^\prime (x - \Delta_n /2)\right \rbrack =0$
 obtained by taking the derivative
of 3.10 at the poles)
$$\eqalign{\left \{ \matrix {
\alpha^2_{\lambda\lambda^\prime}\cr
\beta^2_{\lambda\lambda^\prime}\cr
\alpha^\ast \beta_{\lambda\lambda^\prime}\cr
\alpha \beta^\ast_{\lambda\lambda^\prime}\cr}\right\}=
\theta(\lambda + \lambda^\prime)
\delta(\lambda - \lambda^\prime) \
+\ \sum_n {1 \over 2 \pi \sqrt{\lambda\lambda^\prime}}
\ &\int \! dx  \lbrack e^{+i({\lambda + \lambda^\prime \over 2})
\Delta_n(x) }
 e^{+i(\lambda - \lambda^\prime)
(x-\hat u_0) } \ \times \cr
&\times \
\left ( {\vert \lambda + \lambda^\prime\vert \over 2}
+{\lambda - \lambda^\prime \over 4} \Delta_n^\prime (x)
\right )\rbrack\cr}\eqn\eqIIIxi
$$
where the sum is over the poles $\Delta_n(x)$ defined by 3.10)
such that $i(\lambda + \lambda^\prime)\Delta_n(x)$
has a negative real part with the
exception of the pole at $i\epsilon$ that yields the first term.
The remaining task is to substitute 3.11) into 2.4)
and to perform the indicated integrations and
summation over $n$.  Integration over $(\lambda - \lambda^\prime)$
gives two terms proportional to $2\pi \delta (x-\hat u_0)$
and $2\pi i \delta^\prime (x-\hat u_0)$ respectively,
which in turn makes the integral over $x$
trivial to give (with $\lambda_T = (\lambda + \lambda^\prime)/2$)
$$\eqalign{\langle T_{uu} \rangle_{ren}
&= \sum_n \int_0^\infty {d\lambda_T \over 2 \pi}
e^{-i \lambda_T \Delta_n(\hat u_0)}\
 ( \lambda_T - \lambda_T \Delta_n^{\prime\ 2} /4
- i \Delta_n^{\prime\prime}/4)\cr
&={1 \over 2 \pi}\sum_n \ \
{1 \over (i \Delta_n)^2} + {1 \over 4 }
\left ( { \Delta_n^{\prime\ 2} \over \Delta_n^2}
- {\Delta_n^{\prime\prime} \over \Delta_n}
\right )\cr
&= {1 \over 2 \pi}\sum_n \ \
{1 \over (i \Delta_n)^2} - {1 \over 4 }
(ln\vert \Delta_n \vert )^{\prime\prime}\cr}\eqn\eqIIIxii$$
where we have dropped the pole at $i \epsilon$
 in the $\alpha^2$
contribution.  Were space-time not curved,
this would be the only pole, so it is a
natural and elegant feature of the formalism
that this contribution be removed.

	The contribution on the infinite part of the
semi-circle in the $\Delta$ integration may not vanish.
This must be checked out from case to case.
A familiar example is offered by the calculation
of the energy density of Rindler vacuum wherein
the local modes are Minkowski $\equiv e^{i \lambda u} / \sqrt{4 \pi
\lambda}$
and the Rindler
modes are the non-analytic functions
$\theta (u) u^{i \omega} / \sqrt{4 \pi \omega}$. As
shown in \refmark{\parentani},
this non-analyticity shows up as a singularity in
$\langle T_{\mu \nu} \rangle_{ren}$ on the horizons.

\chapter {Black Hole Evaporation}

	As in Section 2, we work in 1 + 1 dimensions and
schematize the problem by describing evaporation
(break-up of vacuum fluctuations into pairs at the
surface of an incipient black hole) in terms of
vacuum fluctuations of Kruskal $U$ modes in $U$-vacuum.
We first prepare for the subtraction by calculating
the affine parameter, $\hat u$, defined by 2.5, along the
light-like geodesic $v=v_0$.  Introducing the light-like Schwarzschild
coordinates in standard notation $u_S(v_S)=t\pm r^\ast$;
$ r^\ast = r + 2M ln\vert1 -2M/r \vert$, the metric is
$ds^2= (1-2M/r)(dt^2-dr^{\ast\ 2})$,
which in terms of Eddington-Finkelstein coordinates ($v_S$ and $r$)
reads
$ds^2= (1-2M/r)dv_S^2-2 dv_Sdr$
It is then seen that the affine coordinate along the
geodesic $v_S=v_{0S}$ is proportional to $r$.  Eq 2.5) gives
$$\hat u_S = {-2 \left\lbrack r(u_S,v_{0S})-2M\right \rbrack \over
1-2M/r_0}
\eqn\eqIIIIi$$
where $r_0$ is the radial distance of the point $u_{0S}$,$v_{0S}$ at which
$\langle T_{uu} \rangle_{ren}$ is to be calculated and the factor
$-2/(1-2M/r0)$
is the jacobian necessary for ${\partial \hat u_S \over \partial u_S}$ to be
equal to $1$. It is noteworthy that it is this coordinate that plays  a
central
role in the method of Damour and Ruffini \refmark{\damour}  which appeared
immediately after Hawking's first  publications (by asking that the vacuum
state
be stationary and analytic in $r-2M$ at the horizon, these authors pick out
unequivocally the Unruh vacuum).  It plays the same pivotal role in what
follows.

	Our local inertial modes are thus
$e^{-i\lambda \hat u_S ( u_S )} / \sqrt {4 \pi \lambda}$, where
$\hat u_S ( u_S, v_{0S})$ is related
to $u_S$ by inverting $u_S=v_{0S}-2r^\ast$ and expressing
$r^\ast(r)$ in terms of $\hat u_S$.
The physical modes are the Kruskal out-modes
$e^{-i\omega U} / \sqrt {4 \pi \omega}$ with $U$
defined in 2.10), expressible in terms of $\hat u_S$
$$U(\hat u_S)=-e^{-u_S(\hat u_S)/4M}={1 \over 4M} (1-2M/r_0)\hat u_S\
e^{-v_{0S}/4M}\ e^{(1-2M/r_0) \hat u_S /4M}
\eqn\eqIIIIii$$
The function $u(x+\Delta/2)$ in Eq 3.9)
is the function $U(\hat u_S)$ evaluated at $\hat u_S = x_S+\Delta/2$
where as in 3.9) $x_S=(\hat u_{1S} + \hat u_{2S})/2$ ;
$\Delta =\hat u_{1S} - \hat u_{2S}$.
One checks
the validity of
the subsequent
contour integration.  The poles 3.10) are thus solutions of
$$(x_S+ \Delta_{n} /2)\  e^{-{1\over 8M}(1-2M/r_0)\Delta_{n}}
\ -\  (x_S- \Delta_{n} /2)\  e^{+{1\over 8M}(1-2M/r_0)\Delta_{n}}
=0 \eqn\eqIIIIiii$$
The $\Delta_{n}$ are pure imaginary as is seen by
reexpressing 4.3) in the form
$$x_S ={\Delta_{n} \over 2} {
e^{+{1\over 8M}(1-2M/r_0)\Delta_{n}}\ +\
e^{ -{1\over 8M}(1-2M/r_0)\Delta_{n}} \over
e^{ +{1\over 8M}(1-2M/r_0)\Delta_{n}}\ -\
e^{ -{1\over 8M}(1-2M/r_0)\Delta_{n}}}\eqn\eqIIIIiv$$
Indeed, setting the imaginary part of 4.4)
to zero, one finds either $Re(\Delta_{n})=0$ or $Im(\Delta_{n})=0$ and for
the relevant values of $x_S$ (corresponding to $r>0$) $Re(\Delta_{n})=0$.
The equation for $Im(\Delta_{n})\equiv B_n$ is then
$$tan\lbrack {1\over 8M}(1-2M/r_0)B_n\rbrack
 = {B_n \over 2 x_S} = {1\over 8M}(1-2M/r_0){ B_n  \over (1-r/2M)}
\eqn\eqIIIIv$$
where we have replaced $x_S=(\hat u_{1S} + \hat u_{2S})/2$
by its expression as a
function of $r=(r(\hat u_{1S})+r(\hat u_{2S}))/2$ using 4.1).
For $r\rightarrow \infty$,
$B_n \rightarrow (8\pi M)\ n$, whereas for $r\rightarrow 2M$,
the $B_n$'s depart from their Hawking thermal values at
the same time as they are rescaled by the local
conformal factor $(1-2M/r_0)$.  Near the horizon
$B_n = 8 \pi M (1-2M/r_0)^{-1} (n-1/2)$ for $r_0-2M \rightarrow 0$.

	First of all, we wish to point out that these
purely imaginary poles $B_n$ are a typical manifestation
of a thermal-like phenomenon.  They correspond to
multiple windings of the physical propagator when
expressed in terms of the inertial coordinates $\hat u$,
at each point of the space time manifold under
consideration.  It is interesting that the Green's
function is not exactly periodic and therefore the
description in terms of local modes is not  purely thermal.
But as we shall show below, for sufficiently high
frequency where $n=1$ dominates (single winding) the
behavior is pure thermal.

	In Fig. 1 we give a schematic sketch of the
rescaled $B_n$'s ($\equiv \bar B_n$), where
$\bar B_n(r)=B_n(r) {1 \over 8 \pi M} (1-2M/r_0)$ as a function of
$(r-2M)$.
Our final result is thus
$$\eqalign{\langle T_{u_Su_S} \rangle_{ren} &= {1 \over 2 \pi}
\sum_n \ \ {1 \over B_n^2(r)} -{1 \over 4} \left ( {dr \over d
x_S}\right)^2
{d^2 \ ln\ \bar B_n \over dr^2}\cr
&= {T^2(r) \over 2 \pi} \sum_n \ \ {1 \over \bar B_n^2(r)}
- 4 \pi^2 M^2 {d^2 \ ln\ \bar B_n \over dr^2}\cr}\eqn\eqIIIIvi
$$
where $T(r)=T_H (1-2M/r)$, $T_H$ being the Hawking temperature ($=1/8 \pi
M$).
{}From the previous exercise (Eqs. 3.7 and 3.8) this
is a rewrite of 2.12) and it evaluation by analytic
means would entail techniques which would duplicate
this exercise.  This is not the interest of the
present formulation.  Rather, one wishes to dissect
its physical significance in terms of the local modes $\lambda$.

	It is instructive in 4.6) to rewrite $\bar B_n (r) = n + \xi_n(r)$.
  $\xi_n$ thus denote the departure of the poles from their thermal
values. $\xi_n(r)$ as a function
of $r$ varies smoothly for each $n$ from zero at $r=\infty$ to
$-1/2$ at $r=2M$.  Its dependence on $n$ for intermediate
values of $r$ is slight.  Were it not for $\xi_n(r)$,
$\langle T_{u_Su_S} \rangle_{ren}$ would
simply be a thermal flux at temperature $T(r)$.
So the question arises: how much of 4.6) is truly the
result of thermal flux?  In the paragraph that follows
we discuss briefly the relative contributions of $\alpha \beta^\ast$ and
$\beta^2$ to the stress tensor.
We show that for sufficiently large
$r$ there is a lower bound $\lambda_{min}$ above
which the local modes contribute to a pure thermal
flux in that the quantum interference terms $\alpha^\ast \beta$
and $\alpha \beta^\ast$ of Eq. 3.4)
become negligible for such modes.  Furthermore, if $\lambda ,
\lambda^\prime$
are also greater than $T(r)^{-1}$ then the modes are distributed in a
Boltzmann  distribution
at temperature $T(r)$.  For sufficiently large $r$ all modes
are distributed in a Planckian distribution and both
the $\alpha^\ast \beta$ terms and the
${d^2 ln\vert B_n\vert \over dr^2}$ term are negligible.

	The $\beta^2$ contribution to $\langle T_{uu} \rangle$ is readily
evaluated
using 3.11) and 3.4) by integrating over $\lambda - \lambda^\prime$ and
$\lambda + \lambda^\prime$ while restricting $\lambda$ and $\lambda^\prime$
to be positive.
$$\langle T_{uu}(\beta^2)\rangle =\sum_n\ \ \int \!d\delta {i \over 8
\pi^2}
{\Delta_n (\hat u_0 + \delta) \over
\left(\delta^2 - {\Delta_n^2 ((\hat u_0 + \delta)\over 4}\right)^2}
\eqn\eqIIIIvii$$
where the variable of integration in 3.11) has been
translated by $\hat u_0$ and we have not considered the
terms proportional to ${d^2 ln\vert\Delta_n\vert \over d\hat u_0^2}$.
If the $\Delta_n$ do not depend
on $\delta$, then the $\delta$ integral may be performed and
$\langle T_{uu}(\beta^2)\rangle = \langle T_{uu} \rangle_{ren}$,
i.e. the $\alpha^\ast \beta$ term does not contribute.
If $\Delta_n$ varies slowly, it may be expanded in a series
around $\hat u_0$ and the $\delta$ integral then yields
$$\langle T_{uu}(\beta^2)\rangle =\sum_n\ \
{1 \over 2 \pi (i \Delta_n)^2} + A {\Delta_n^{\prime 2} \over \Delta_n^2}
+B{ \Delta_n^{\prime\prime} \over \Delta_n}
\eqn\eqIIIIviii$$
where $A$ and $B$ are uninteresting numerical coefficients.
Far from the black hole $\bar B_n =n\pi - 2 M n\pi /r +O(1/r^2)$
and both the $(ln\vert\bar B_n\vert )^{\prime\prime}$  and the
$\alpha^\ast \beta$ contribution
vanishes as $1/r^3$.

	More interesting though is the behavior of $\alpha^2$,$\beta^2$,
$\alpha^\ast \beta$ for large $\lambda$,$\lambda^\prime$.
Unfortunately a straightforward analysis is not forthecoming in the black
hole
case due to the negative curvature of $\bar B_n(r)$ (see figure 1). We have
attempted to circumvent this difficulty by changing the basis functions
from
plane waves to wave packets and have shown that at large distance from the
black hole when $\lambda , \lambda^\prime > B^{\prime\prime}_n(r) =
{32M^2n\pi \over r^3}$ the contribution of the n'th pole to $\alpha^\ast
\beta$
vanishes exponentially. Near the black hole we have not yet been able to
reach
definite conclusions.

\ack
The authors wish to thank Philippe Spindel for stimulating discussions.

\refout
\vskip 2 cm
\centerline{Figure Caption}

Figure 1. The thermal poles for $\langle T_{u_Su_S} \rangle_{ren}$ around
an
evaporating black hole as a function of r. The $\bar B_n$'s are
plotted as function of $(r-2M)/2M$ , where
$\bar B_n (r) = B_n (r) {1 \over 8 \pi M}(1-2M/r)$ and
$B_n$ is given by equation 4.5)

\end